\documentclass{PoS}
\usepackage{hepnames}
\usepackage{ragged2e}
\usepackage{hepunits}
\newcommand{\fm}{\mathop{\rm fm}\nolimits}

\newcommand{\rd}[1]{\mathop{\mathrm{d}#1}}

\newcommand{\deriv}[2]{\frac{\mathrm{d}#1}{\mathrm{d}#2}}

\newcommand{\pderivv}[1]{\frac{\partial}{\partial#1}}

\newcommand{\inv}[1]{\frac{1}{#1}}

\newcommand{\ten}[1]{10^{#1}}

\newcommand{\cd}[2]{\nabla_{#1}{#2}}

\DeclareMathOperator\arctanh{arctanh}
\title{Evolution of magnetic fields in a transversely expanding highly conductive fluid}

\ShortTitle{Evolution of magnetic fields in a transversely expanding highly conductive fluid}
\author{\speaker{Masoud~Shokri} and N$\acute{\mbox{e}}$da~Sadooghi\\
        Department of Physics,
        Sharif University of Technology,
P.O. Box 11155-9161, Tehran, Iran\\
        E-mail: \email{m\_shokri@physics.sharif.ir},
    \email{sadooghi@physics.sharif.ir}}
\abstract{Due to the absence of a transverse expansion with respect to the beam direction, the Bjorken flow is unable to describe certain observables in heavy ion collisions. This caveat has motivated the introduction of analytical relativistic hydrodynamics (RH) solutions with transverse expansion, in particular, the 3+1 self-similar (SSF) and Gubser flows. Inspired by recent generalizations of the Bjorken flow to the relativistic magnetohydrodynamics (RMHD), we present a procedure for a generalization of RH solutions to RMHD. Our method is mainly based on symmetry arguments. Using this method, we find the relation between RH degrees of freedom and the magnetic field evolution in the ideal limit for an infinitely conductive fluid, and determine the proper time dependence of the magnetic field in aforementioned flows. In the case of SSF, a family of solutions are found that are related through a certain differential equation. To find the magnetic field evolution in the Gubser flow, we solve RMHD equations for a stationary fluid in a conformally flat $dS^3\times E^1$ spacetime. The result is then Weyl transformed back into the Minkowski spacetime. In this case, the temporal evolution of the magnetic field exhibits a transmission between $1/t$ to $1/t^3$ near the center of the collision. The longitudinal component of the magnetic field is found to be sensitive to the transverse size of the fluid. We also find the radial evolution of the magnetic field for both flows. The radial domain of validity in the case of SSF is highly restricted, in contrast to the Gubser flow. A comparison of the results suggests that the Gubser RMHD may give a more appropriate qualitative picture of the magnetic field decay in the quark-gluon plasma (QGP).}

\FullConference{XIII Quark Confinement and the Hadron Spectrum - Confinement2018\\
		31 July - 6 August 2018\\
		Maynooth University, Ireland}

\begin{document}

\section{Introduction}
In heavy ion colliders, heavy nuclei (gold at the RHIC and lead at the LHC) are accelerated up to ultra-relativistic energies, and eventually collided in order to produce a hot medium. Since the nuclei comprise fast moving positive charges, large electromagnetic fields are generated during the early stages of heavy ion collisions. The magnetic field at top RHIC energies, i.e. $\sqrt{s_{NN}}=200~\GeV$, is estimated to be in the order of $\ten{18}$ Gauss, and at LHC energies it can reach even higher orders of magnitude \cite{warringa2007,skokov2009,Huang-review}. The main source of the electromagnetic fields are the nucleons that do not participate in collisions. They are referred to as spectators. They quickly leave the scene, and if there were no medium produced, electromagnetic fields would die as the sources leave the scene. However, as it turns out, at some short times after the collision a plasma of quarks and gluons is formed that mutually evolves with the fields, and may slow down their decay. Relativistic magnetohydrodynamics provides an effective picture of the mutual evolution of the QGP and electromagnetic fields.
\par
The constituent equations of RMHD are the energy-momentum conservation and Maxwell equations
\begin{equation}
	\cd{\mu}{T^{\mu\nu}}=0,\quad\cd{\nu}{F^{\mu\nu}}=J^\mu,\quad\epsilon^{\mu\nu\alpha\beta}\cd{\nu}{F_{\alpha\beta}}=0.
\end{equation}
Here, $T^{\mu\nu}$, $F^{\mu\nu}$ and $J^\mu$ are the energy-momentum tensor, field strength tensor and electric current four-vector. Moreover, $\epsilon$ is the energy density of the fluid.  If the fluid is highly conductive, one may consider the ideal magnetohydrodynamic (MHD) limit
\begin{equation}
u_\mu F^{\mu\nu} = 0,
\end{equation}
where $u$ is the fluid four-velocity. In this approximation, the local electric field is zero (infinitely small), and the electromagnetic degrees of freedom are encoded in the magnetic four-vector
\begin{equation}
B^\mu=\inv{2}\epsilon^{\mu\nu\rho\sigma}u_\rho F_{\rho\sigma}.
\end{equation}
To the best of our knowledge, the first generalization of an analytical solution of relativistic hydrodynamics to RMHD, in the QGP context, is presented in \cite{Rischke-bjorken-mhd}. Here, the Bjorken flow \cite{bjorken} is considered in the ideal MHD limit. In particular, it is found that the magnetic field evolves similar to an arbitrary conserved charge density, $Q$, being the solution to the continuity equation
\begin{equation}
\cd{\alpha}{\left(Qu^\alpha\right)}=0.
\end{equation}
Physical examples of such a charge density are the baryon number density in a baryonic fluid or the entropy density in a perfect fluid. For our purpose, $Q$ does not need to be any physical quantity. To determine the evolution of the magnetic field in an ideal magnetized fluid, it is appropriate to reparameterize the Minkowski spacetime as follows
 \begin{equation}
 {\rd{s}}^2 = -{\rd{\tau}}^2+r^2{\rd{\phi}}^2+\tau^2{\rd{\eta}}^2+{\rd{r}}^2,
 \end{equation}
where
 \begin{equation}
 \tau=\sqrt{t^2-z^2},\quad\eta=\arctanh(z/t),\quad \phi=\arctan(y/x),\quad r=\sqrt{x^2+y^2}.
 \end{equation}
The evolution of $Q$ and the magnitude of the local magnetic field, i.e. $B=\sqrt{B^\mu B_\mu}$, in this case are found to be
  \begin{equation}
  Q=Q_0\frac{\tau_0}{\tau},\quad\mbox{and}\quad B=B_0\frac{\tau_0}{\tau}.
  \end{equation}
In the Bjorken flow, the fluid doesn't expand in transverse directions with respect to the beam direction. However, it is known that certain final observables of heavy ion collisions cannot be understood without considering a significant transverse expansion \cite{CasalderreySolana:2011us}. This fact motivated generalizations of the Bjorken flow to flows with transverse expansion. Two examples of such flows are the 3+1 dimensional self-similar \cite{Csorgo-cyl} and Gubser flows \cite{Gubser-Symmetry,Gubser-Conformal}. In \cite{shokri-JHEP}, we have shown that in both cases the magnetic field is related to $Q$. However, the relationship between $Q$ and the magnetic field is more complicated than in the Bjorken case. In what follows, we study the generalization of aforementioned flows in RMHD and, in particular, the effects of transverse expansion on the evolution of magnetic fields.
\section{Symmetry arguments in RH and RMHD}
\subsection{Symmetries in RMHD}	
To find the evolution of magnetic fields, we use appropriate symmetry arguments. Let us  first consider the pure hydrodynamic case. The four-velocity of a fluid comprises three independent components. Thus, putting three independent symmetry constraints may fix the four-velocity. Let us denote such a set of symmetries by $\mathcal{I}$ that includes at least three independent Killing vectors
\begin{equation}
\mathcal{I}=\{\xi_1,\xi_2,\xi_3,\cdots\}.
\end{equation}
If $\Gamma$ is a scalar that respects all of the symmetries in $\mathcal{I}$, then $u_\mu$ may be proportional to $\Gamma$'s derivatives. Hence, with a fixed four-velocity, we are, in principle, able to find the arbitrary conserved charge density. However, for a RMHD generalization, we find it more appropriate to relax some of the symmetries in $\mathcal{I}$, and introduce a proper similarity variable $\vartheta$,\footnote{This means that $\vartheta$ satisfies $u^\mu\partial_\mu{\vartheta}=0$.} that respects the remained symmetries in a way that the same velocity profile is fixed. Let us call this subset of symmetries $\mathcal{M}$ which comprises two Killing vectors. Since the similarity variable does not change through flow lines, it can be used to label them.
For simplicity, we transform to a coordinate system in which the assumed symmetries in $\mathcal{M}$ imply that nothing depends on coordinates $x_1$ and $x_2$. Applying these symmetries to the homogeneous Maxwell equations in the ideal MHD limit, we find the relations between the field strength tensor and the arbitrary conserved charge density as
\begin{equation}
F_{13}=Q(x^0,x^3)u^0\sqrt{-g}f(\vartheta)h(\vartheta),\quad
F_{23}=Q(x^0,x^3)u^0\sqrt{-g}h(\vartheta),\quad
F_{12}=0.
\end{equation}
In order to find the scaling functions $f(\vartheta)$ and $h(\vartheta)$, we have to solve the Euler equation
\begin{equation}\label{euler}
\left(\epsilon+p+B^2\right)a^\mu = -\Delta^{\mu\nu}\left[\partial_\nu\left(p+\frac{B^2}{2}\right)-\cd{\rho}{\left(B_\nu B^\rho\right)}\right].
\end{equation}
Here, $a^\mu=u^\nu\cd{\nu}{u^\mu}$ is the fluid proper acceleration, and $\Delta^{\mu\nu}=g^{\mu\nu}+u^\mu u^\nu$ projects four-vectors to the plane perpendicular to the four-velocity.
In both cases of 3+1 dimensional SSF and Gubser flows, we assume that RMHD is invariant under boost along and rotation around the beamline
\begin{equation}
\mathcal{M}=\{\pderivv{\phi},\pderivv{\eta}\}.
\end{equation}
By this assumption, we are dealing with near-central collisions in the mid-rapidity region.
\subsection{Magnetic field in SSF}
The SSF is a combination of three simultaneous Bjorken flows in three spatial directions. The symmetry group of SSF is the Lorentz group, and its invariant scalar is $\varrho\equiv\sqrt{\tau^2-r^2}$. We choose the proper similarity variable to be $\mathbf{\vartheta=r/\tau}$. The Euler equation \eqref{euler} thus reduces to one equation for two unknowns
\begin{equation}
\inv{2}\left(1-\vartheta^2\right)\left[(\vartheta^2+\mathcal{A}_2^2\mathcal{F}(\vartheta))\deriv{\mathcal{H}}{\vartheta}+\mathcal{A}_2^2\mathcal{H}(\vartheta)\deriv{\mathcal{F}}{\vartheta}\right]
+2\vartheta\mathcal{H}(\vartheta)\left(1+\mathcal{A}_2^2\mathcal{F}(\vartheta)\right)=0.
\end{equation}
We have redefined the scaling functions for simplicity
\begin{equation*}
\mathbf{\mathcal{A}_1\sqrt{\mathcal{H}(\vartheta)}\equiv Q_0h(\vartheta),\qquad \mathcal{A}_2\sqrt{\mathcal{F}(\vartheta)}\equiv -f(\vartheta).}
\end{equation*}
Here $\mathcal{A}_1$ and $\mathcal{A}_2$ are constants. 
There is no unique solution for the magnetic field. A specific solution is, nevertheless, found if we assume the electric current to be zero. It is given by
\begin{equation}\label{ssf-b}
B=\frac{B_0}{\sqrt{1+\alpha_0^2	}}\left(\frac{\varrho_0}{\varrho}\right)^2\sqrt{\left(\frac{r_0}{r}\right)^2+\alpha_0^2\left(\frac{\tau_0}{\tau}\right)^2}.
 \end{equation}
Here, $B_0$ is the magnitude of the magnetic field and $\alpha_0$ is the ratio $B_z/B_y$, both at the initial time and radius $\tau_{0}$ and $r_{0}$.
\subsection{Conformal MHD}
There are two approaches to the Gubser flow. The first one uses special conformal transformations in the flat spacetime. In this approach, the symmetry group of the Gubser flow is found by replacing the translational invariance in the transverse plane with the  invariance under a certain special conformal transformation \cite{Gubser-Symmetry}. The invariant scalar,
\begin{equation}\label{gubser-invariant}
G\equiv\frac{1-q^2\left(\tau^2-r^2\right)}{2q\tau},
\end{equation}
contains the inverse of the typical transverse size of the fluid, $q$. As it turns out, one cannot introduce a similarity variable in this approach. Moreover, the full set of Gubser symmetries eliminates the transverse components of the magnetic field \cite{shokri-JHEP}. This is, however, not a feature of heavy ion collisions. The second approach is to solve hydrodynamics equations in a conformally flat spacetime, and then transform the solutions back to the flat spacetime \cite{Gubser-Conformal}. A conformally flat spacetime, described by the metric ${\rd{\hat{s}}}^2$, is found by a Weyl transformation that is multiplying the flat spacetime metric, ${\rd{s}}^2$, by a local function $\Omega(x)$,
\begin{equation}\label{weyl}
{\rd{s}}^2=\Omega(x)^2{\rd{\hat{s}}}^2.
\end{equation}
We use this second approach to find the magnetic field evolution in the Gubser flow. The essence of this approach is that a stationary fluid in a conformally flat spacetime may transform into an accelerated fluid in the flat spacetime. In particular, a stationary fluid in $dS^3\times E$ corresponds to the Gubser flow in the flat spacetime, provided  $\Omega$ is set equal to $\tau$ in \eqref{weyl}. To determine the magnetic field evolution in this setup, we parameterize $dS^3\times E$ metric as
\begin{equation}
{\rd{\hat{s}}}^2=-{\rd{\rho}}^2+\cosh^{2}\rho~\sin^{2}\theta~{\rd{\phi}}^2+{\rd{\eta}}^2+\cosh^{2}\rho~{\rd{\theta}}^2,
\end{equation}
with
\begin{equation}
\sinh\rho=-\frac{1-q^2\tau^2+q^2r^2}{2q\tau},\quad\mbox{and}\quad
\tan\theta=\frac{2qr}{1+q^2\tau^2-q^2r^2}.
\end{equation}
We further assume $\rho$ as the invariant scalar and $\theta$ as the proper similarity variable. We apply the symmetries in $dS^3\times E$, and solve the Euler equation \eqref{euler}. In this way, the magnetic field in $dS^3\times E$ is found. We only need to transform the result back into the Minkowski spacetime. The final result reads (see \cite{shokri-JHEP} for more details)
\begin{equation}\label{gubser-b}
B=\mathcal{A}_1\sqrt{\frac{1}{r^2\tau^2}+\beta_0^2\left(\frac{4q^2}{1+q^4(\tau^2-r^2)^2+2q^2(\tau^2+r^2)}\right)^2}.
\end{equation}
It turns out that the magnitude of the magnetic field comprises a Bjorken-like part from the transverse component and a size-dependent part from the longitudinal component.
The free parameters $\mathcal{A}_1$ and $\beta_0$ are fixed by initial conditions.
If the longitudinal component is initially absent, or the transverse size of the fluid is very large, the evolution of magnetic field at any fixed radius is similar to the Bjorken MHD. Interestingly the electric current automatically vanishes in this case.
\subsection{Comparison of the results}
\begin{figure}[hbt]
\begin{center}
		\includegraphics[width=12cm,height=8cm]{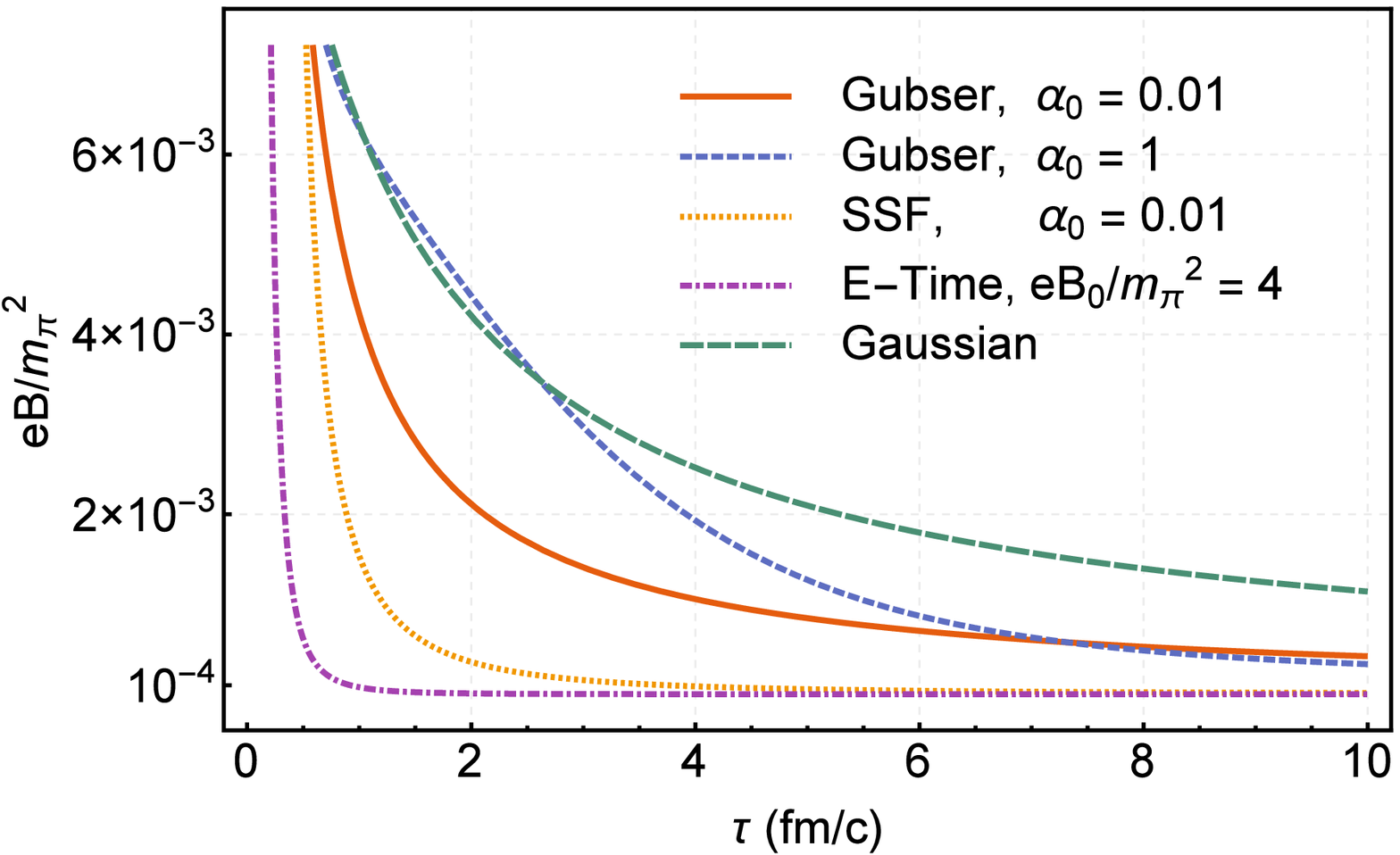}
		\caption{The temporal evolution of \ensuremath{eB/m_{\pi}^{2}} for our solutions is compared. Here, the pion mass \ensuremath{m_{\pi}=137} MeV. The red solid and blue dashed curves show the Gubser MHD solution \protect\eqref{gubser-b} with weak and strong longitudinal components, characterized by small and large \ensuremath{\alpha_{0}=0.01,1}. The yellow dotted curve is the SSF solution \protect\eqref{ssf-b} with a weak longitudinal component with \ensuremath{\alpha_{0}=0.01}. It turns out that the SSF solution is not significantly altered by a change in the longitudinal component. The purple dotted-dashed curve is the early-time (E-time) dynamics \protect\eqref{early-time} with \ensuremath{eB_{0}/m_{\pi}^{2}=4}. The green dashed curve arises from the Gaussian ansatz \protect\eqref{gaussian}. The SSF solution is very similar to the early-time dynamics, and exhibits a very fast \ensuremath{\tau^{-3}} decay. On the other hand, Gubser MHD shows a transition from a Bjorken-like decay, i.e. \ensuremath{\tau^{-1}} in early times to a \ensuremath{\tau^{-3}} in later times.}\label{fig-all-pion-tau}
\end{center}
\end{figure}

In Fig. \ref{fig-all-pion-tau}, we compare our SSF and Gubser solutions for $B$, \eqref{ssf-b} and \eqref{gubser-b}, with the early time dynamics
\begin{equation}\label{early-time}
eB(\tau)=\frac{eB(0)}{\left(1+\tau^2/t_B^2\right)^{3/2}},
\end{equation}
proposed by \cite{Huang-review} and the Gaussian ansatz
\begin{equation}\label{gaussian}
\frac{eB(\tau,r)}{m_{\pi}^2} = \inv{a_1+b_1\tau}\exp\left(-\frac{r^2}{\sigma_r^2}\right),
\end{equation}
presented in \cite{Roy-reduced-mhd}, and based on a numerical solution from \cite{Zakharov}.
In \eqref{early-time}, $t_B$ is the typical time scale that spectators pass the collision stage and $eB(0)$ is the maximum magnetic field reached in the collision. In the Gaussian ansatz \eqref{gaussian}, $a_1=78.2658$, $b_1=79.5457\fm^{-1}$ and $\sigma_r=3.5\fm$ for a central collision.
\section{Conclusion}
Inspired by Bekenstein and Oron's work, initially done in a different context \cite{Bekenstein-cons-laws}, we presented a procedure for determining the dynamics of magnetic fields in the presence of a highly conductive fluid that expands with a given velocity profile. This procedure is based on the assumption of two symmetries and one proper similarity variable. Introducing a similarity variable seems to be crucial, in particular, in the RMHD generalization. Despite the theoretical nature of our work, it may provide some insights into the evolution of magnetic fields in heavy ion collisions at the QGP phase. Following conclusion are thus made:
\begin{enumerate}
\item Even with an early transverse expansion, the decay of the magnetic field in the QGP stage is significantly slower than in the pre-equilibrium gluon dominated stage.\vspace{-0.2cm}
\item Due to the finite size of the fluid, a nonvanishing longitudinal magnetic field may play an unexplored role in heavy ion collisions. We find that even a small initial longitudinal component may enhance during the expansion of the fluid.\vspace{-0.2cm}
\item
The ratio of the magnetic field energy density to the fluid energy density may also enhance during the expansion of the fluid.
\end{enumerate}
\begin{justify}
We close by mentioning two directions in which this work can be extended.
One direction is to go beyond the ideal MHD limit. Although the conductivity of the QGP is very large in comparison with ordinary matter, its magnetic Reynolds number may not be as large as required for the ideal MHD limit to be valid \cite{Huang-review}. This is a motivation for studying RMHD solutions for finite conductivities, as we previously did for Bjorken MHD \cite{rotating}. Another direction is to consider the noncentral collisions in which rotational invariance is a poor approximation, and electromagnetic fields are even more significant. However, an analytical treatment of noncentral collisions is much more difficult. Recently, an analytical solution to RH concerning noncentral collisions was introduced in \cite{Bantilan:2018vjv}. We postpone the generalization of this flow to RMHD to our future publications.
\end{justify}

\end{document}